\documentclass[aps,prb,superscriptaddress,citeautoscript,twocolumn]{revtex4-1}

\usepackage{graphicx}
\usepackage{amsmath}
\usepackage{amssymb}
\usepackage{color}

\newcommand{\bea}{\begin{eqnarray*}}
\newcommand{\eea}{\end{eqnarray*}}
\newcommand{\bne}{\begin{equation*}}
\newcommand{\ede}{\end{equation*}}

\newcommand{\bnen}{\begin{equation}}
\newcommand{\eden}{\end{equation}}
\newcommand{\bean}{\begin{eqnarray}}
\newcommand{\eean}{\end{eqnarray}}
\newcommand{\bnsn}{\begin{subequations}}
\newcommand{\edsn}{\end{subequations}}

\newcommand{\bna}{\begin{array}}
\newcommand{\eda}{\end{array}}
\newcommand{\bnm}{\begin{enumerate}}
\newcommand{\edm}{\end{enumerate}}

\newcommand{\abs}[1]{\left| #1 \right|}

\newcommand{\ket}[1]{| #1 \rangle}
\newcommand{\bra}[1]{\langle #1 |}

\begin{document}

\title{Transport signatures of an Andreev molecule in a quantum dot -- superconductor -- quantum dot setup}
\author{Zolt\'an Scher\"ubl}
%\email{scherubl.zoltan@gmail.com}
\affiliation{Department of Physics and MTA-BME Momentum Nanoelectronics Research Group, Budapest University of Technology and Economics, Budafoki \'ut 8., 1111 Budapest, Hungary}
\author{Andr\'as P\'alyi}
\affiliation{Department of Theoretical Physics and MTA-BME Exotic Quantum Phases "Momentum" Research Group, Budapest University of Technology and Economics, 1111 Budapest, Hungary}
\author{Szabolcs Csonka}
\affiliation{Department of Physics and MTA-BME Momentum Nanoelectronics Research Group, Budapest University of Technology and Economics, Budafoki \'ut 8., 1111 Budapest, Hungary}
%\email{csonka@dept.phy.bme.hu}

\begin{abstract}
Hybrid devices combining quantum dots with superconductors are important building blocks of conventional and topological quantum-information experiments. A requirement for the success of such experiments is to understand the various tunneling-induced non-local interaction mechanisms, namely, crossed Andreev reflection, elastic cotunneling, and direct interdot tunneling, that are present in the device. Here, we provide a theoretical study of a simple device which consists of two quantum dots and a superconductor tunnel-coupled to the dots, often called a Cooper-pair splitter. We study the three special cases where one of the three non-local mechanisms dominates, and calculate measurable ground-state properties, as well as the zero-bias and finite-bias differential conductance characterizing electron transport through this device. We describe how each non-local mechanism controls the measurable quantities, and thereby find experimental fingerprints that allow one to identify and quantify the dominant non-local
mechanism using experimental data. Finally, we study the triplet blockade effect and the associated negative differential conductance in the Cooper-pair splitter, and show that they can arise regardless of the nature of the dominant non-local coupling mechanism. Our results should facilitate the characterization of hybrid devices, and their optimization for various quantum-information-related experiments and applications. 
\end{abstract}

\date{\today}
\maketitle

\section{Introduction}

Superconducting hybrid nanodevices provide a promising platform for quantum architecture. While superconductors (SCs) allow for a spatially extended coherent state, nanodevices provide the confinement of electrons into 1D or 0D. The interplay of these properties is a key ingredient of novel promising qubit realizations, such as Majorana qubits \cite{Leijnse_review} and Andreev qubits \cite{JanvierScience2015}. 

The basic physical mechanism behind these applications is the Andreev reflection, when a Cooper pair from the SC is transformed to two electrons in the normal conductor. The conversion of Cooper pairs has a special form called crossed Andreev reflection (CAR), when the two electrons originating from the Cooper pair ends up in spatially separated normal parts \cite{ByersPRL1995,DeutscherAPL2000,LesovikEurPhysJB2001,RecherPRB2001}. 

CAR is also a potential resource for quantum hardwares. On the one hand, it naturally generates entangled spatially separated electron pairs \cite{RecherPRB2001}. On the other hand, several novel topological superconducting proposals are based on CAR process, such as the poor man's Majorana setup \cite{LeijnsePRB2012}, the Majorana chain \cite{SauNatComm2012}, Tritops \cite{KlinovajaPRB2014_2,HaimPRB2014,GaidamauskasPRL2014,KeselmanPRL2013,LiuPRX2014}, Majorana states in graphene \cite{ClarkeNatPhys2014,SanJosePRX2015,LeeNatPhys2017} and devices with even more exotic non-Abelian excitations, such as parafermions \cite{ClarkeNatComm2013,KlinovajaPRB2014,AliceaAnnRevCondMatPhys2016}. CAR was studied experimentally in metallic nanostructures \cite{BeckmannPRL2004,RussoPRL2005,Cadden-ZimanskyPRL2006,BeckmannApplPhysA2007} and later in so-called Cooper pair splitter devices, where two quantum dots (QDs) are weakly tunnel coupled to a superconductor in a QD-SC-QD geometry \cite{HofstetterNature2009,HermannPRL2010,HofstetterPRL2011,DasNatComm2012,WeiNatPhys2010,SchindelePRL2012,LambertPRB2014,TanPRL2015}. The QD-SC-QD setup serves also as the basic building block of the poor man's Majorana setup \cite{LeijnsePRB2012} and the Majorana chain \cite{SauNatComm2012}.

Strong tunnel coupling between a QD and a SC leads to the formation of Andreev bound states (ABSs) \cite{JespersenPRL2007,GovernalePRB2008,MengPRB2009,DeaconPRL2010,DeaconPRB2010,PilletNatPhys2010,FuttererPRB2013,KimPRL2013,ChangPRL2013,PilletPRB2013,KumarPRB2014,SchindelePRB2014,LeeNatNano2014,JellinggaardPRB2016,LeePRB2017,GramichPRB2017,LiPRB2017,ZitkoPRB2015,BretheauNatPhys2017} via local Andreev reflection (LAR). Due to the charging energy on the QDs, the QD-SC-QD geometry prefers CAR process over the LAR and leads to the expectation that CAR hybridizes the states of the two QDs, generating the so-called Andreev molecular state \cite{EldridgePRB2010,TrochaPRB2015,WrzesniewskiJPCM2017}. The first experimental realization towards such state is already reported \cite{SuNatComm2017}. However, CAR is not the only coupling mechanism between the QDs. Electrons can be transmitted from one of the QDs to the other via virtual intermediate SC quasiparticle states, via the so-called elastic cotunneling (EC) process \cite{AverinPRL1990,FalciEuroPhysLett2001,MelinPRB2004,HasslerPRB2015}. Furthermore, if there is direct tunnel coupling between the dots, as in certain experimental realizations \cite{FulopPRB2014,FulopPRL2015}, then this interdot coupling (IT) also influences the spectrum and the dynamics. 

In an experimental realization of a QD-SC-QD setup, any of the three non-local coupling mechanisms (CAR, EC or IT) could dominate. The focus of this paper is to calculate measurable quantities and explore differences between the individual fingerprints of the three non-local processes, which allow to identify and quantify the dominant non-local term. In particular, we describe the ground-state properties (phase diagram, average electron occupation) of the system, the zero-bias conductance describing electron transport through the device in the presence of tunnel-coupled normal leads, and the excitation spectrum that is accessible via finite-bias transport measurements. 

The paper is organized as follows: First, we introduce the microscopic model of the Cooper-pair splitter, and outline how to derive a simple effective Hamiltonian, which describes the non-local coupling mechanisms as individual Hamiltonian terms, from the microscopic model. Then, we analyze the ground state properties, the phase diagram and average electron occupation, of the QD-SC-QD system. Subsequently, we outline the transport model describing the setup where each QD is coupled to a normal lead. Finally, we analyze the transport signatures of the different non-local coupling terms via the zero bias conductance, the finite-bias conductance, and the triplet blockade effect.

\section{Model}

\subsection{Microscopic Hamiltonian of the proximitized double quantum dot}

Throughout this work, we study a standard Cooper pair splitter device. The setup is shown in Fig.~1a. It consists of two QDs, each of them tunnel-coupled to its own normal (N) lead, and a common superconducting lead  (SC). The Hamiltonian of the system is:
\bnen H = H_{\text{QD}}+ H_{\text{SC}} + H_{\text{T,SC}} + H_{\text{IT}}+ H_{\text{N}} + H_{\text{T,N}}. \eden
We assume that the level spacings of the dots are large, i.e. each QD has a single spinful orbital, which can be occupied by 0, 1 or 2 electrons. The first term in the Hamiltonian reads as
\bnen H_{\text{QD}} = \sum_{\alpha=L,R} \left ( \sum_{\sigma=\uparrow,\downarrow} \varepsilon_{\alpha} d^{\dagger}_{\alpha\sigma} d_{\alpha\sigma} + U_\alpha n_{\alpha\uparrow} n_{\alpha\downarrow} \right), \eden
where $\varepsilon_{\alpha}$ is the on-site energy of QD$_{\alpha}$ ($\alpha=L,R$), $d_{\alpha\sigma}$ ($d^{\dagger}_{\alpha\sigma}$) annihilates (creates) an electron on QD$_{\alpha}$ with spin $\sigma$, and $U_\alpha$ is the on-site Coulomb repulsion energy. Note that the interdot Coulomb repulsion in neglected here, since the SC lead between the QDs screens this interaction. Throughout this work, we assume identical Coulomb repulsion energies in the two dots, and use this energy scale $U = U_L = U_R$ as the unit of energy. 

\begin{figure*}
\begin{center}
\includegraphics[width=17cm,keepaspectratio]{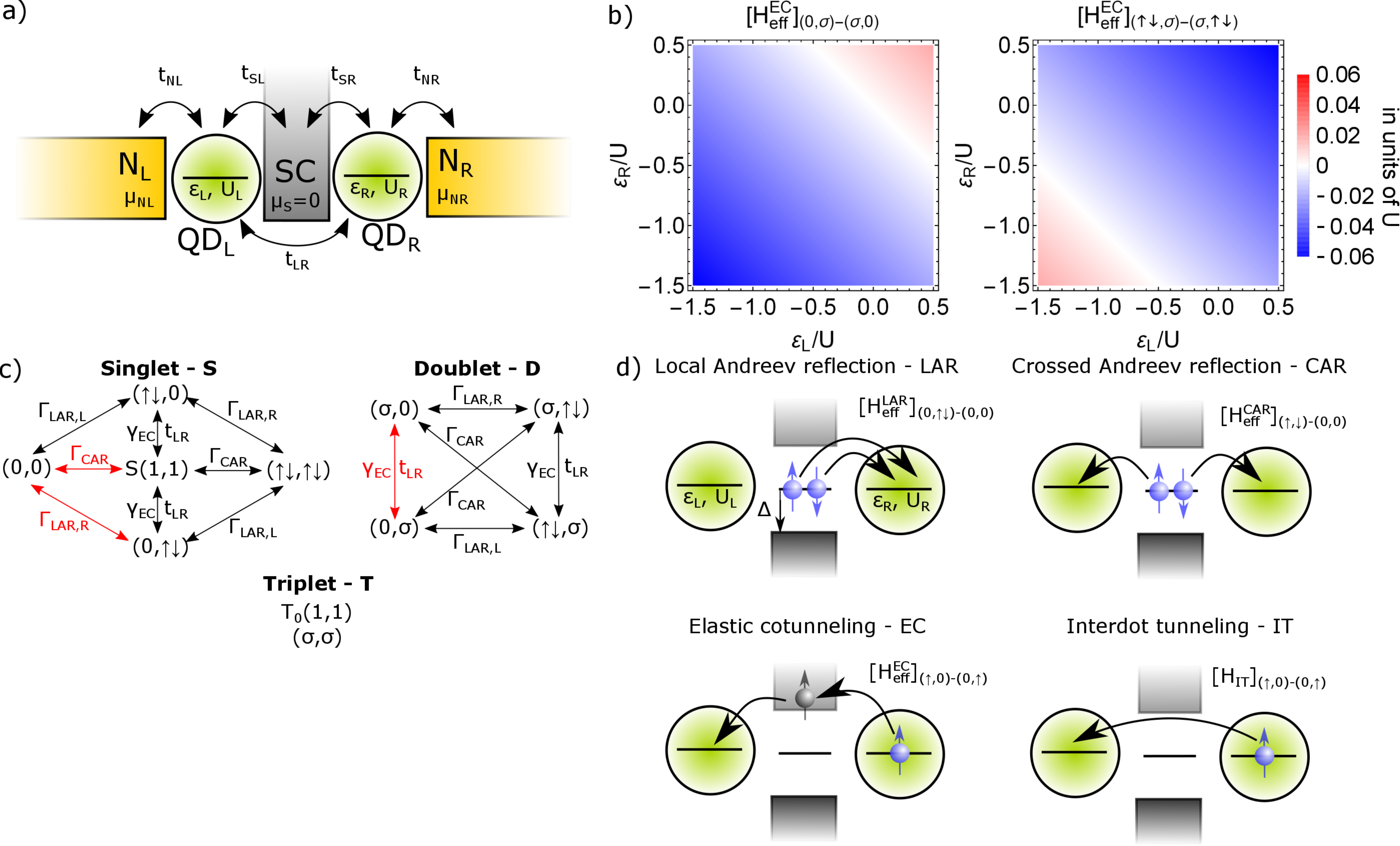}
\caption{
Coherent hybridization between a superconductor and two quantum dots in a Cooper pair splitter. 
a) Schematics of the Cooper pair splitter setup. 
b) Two examples of the dependence of the EC matrix elements on the on-site energies of the QDs.
c) Invariant subspaces of the effective Hamiltonian $H_{\text{eff}}$ describing the QD-SC-QD setup, also showing the couplings between the basis states. 
d) Examples of the coupling processes encoded in the effective Hamiltonian $H_\text{eff}$. LAR couples, e.g., state $\ket{0,0}$ to state $\ket{0,\uparrow\downarrow}$ by transferring a Cooper pair from SC to QD$_{\text{R}}$. CAR splits a Cooper pair by filling both QDs with one electron with opposite spins. EC transfers an electron from one QD to the other via a virtual intermediate quasiparticle state in SC. IT transfers an electron from one QD to the other without any interaction with SC.}
\label{fig1}
\end{center}
\end{figure*}

The SC lead is described by the standard mean-field Bardeen-Cooper-Schrieffer (BCS) Hamiltonian
\bnen H_{\text{SC}} = \sum_{k\sigma} \varepsilon_{Sk} c^{\dagger}_{Sk\sigma} c_{Sk\sigma} - \Delta \sum_k \left( c_{S-k\downarrow} c_{Sk\uparrow} + h.c. \right), \eden
where $\varepsilon_{Sk}$ is the dispersion of conduction electrons in the SC, $c_{Sk\sigma}$ ($c^{\dagger}_{Sk\sigma}$) annihilates (creates) an electron in the SC with momentum $k$ and spin $\sigma$, and $\Delta$ is the superconducting order parameter. Only one SC lead is present in our setup, hence $\Delta$ is chosen to be real. $H_{\text{SC}}$ can be diagonalized by a Bogoliubov-transformation, $c_{Sk\sigma}=u_k \gamma_{k\sigma} + \sigma v_k \gamma^{\dagger}_{-k\bar{\sigma}}$, where
\bnen \label{eq:BogAmp} \begin{pmatrix} u_k \\ v_k \end{pmatrix} = \left[\frac{1}{2} \left(1 \pm \frac{\varepsilon_{Sk}}{E_k} \right) \right]^{1/2}, \eden
resulting in
\bnen H_{\text{SC}} = \sum_{k\sigma} E_k \gamma^{\dagger}_{k\sigma} \gamma_{k\sigma}, \eden
where $E_k=\sqrt{\varepsilon^2_{Sk}+\Delta^2}$ the quasi-particle energy.

The Hamiltonian of the normal leads is 
\bnen H_{\text{N}} = \sum_{\alpha k\sigma} \varepsilon_{\alpha k} c^{\dagger}_{\alpha k\sigma} c_{\alpha k\sigma}, \eden
where $\alpha = L, R$ is the lead index, $\varepsilon_{\alpha k}$ is the dispersion relation, and $c_{\alpha k \sigma}$ $\left(c^{\dagger}_{\alpha k \sigma} \right)$ annihilates (creates) an electron with momentum $k$ and  spin $\sigma$ in lead N$_\alpha$.

Tunneling between the three leads and the two dots is described by the following terms: 
\bean H_{\text{T,SC}} &=& \sum_{\alpha k\sigma} \left( t_{S\alpha} c^{\dagger}_{Sk\sigma} d_{\alpha\sigma} + h.c. \right), \nonumber \\ H_{\text{T,N}}  &=& \sum_{\alpha k\sigma} \left( t_{N\alpha} c^{\dagger}_{\alpha k\sigma} d_{\alpha\sigma} + h.c. \right), \eean
where $H_{\text{T,SC}}$ describes the tunneling between the QDs and the SC lead, while $H_{\text{T,N}}$ between the dots and the N leads, with $t_{S\alpha}$ ($t_{N\alpha}$) being the tunneling amplitude between the SC (N$_{\alpha}$) lead and QD$_{\alpha}$. Tunneling to the SC will be treated coherently, while tunneling to the N leads is assumed to be weak, and are treated by Fermi's Golden Rule in the transport model outlined below.

Finally,
\bnen H_{\text{IT}} = t_{\text{LR}}  \sum_{\sigma} \left( d^{\dagger}_{L\sigma} d_{R\sigma} + h.c. \right) \eden
describes interdot tunneling (IT), i.e., direct tunneling between the QDs, with an amplitude  $t_{\text{LR}}$.

\subsection{Effective Hamiltonian of the proximitized double quantum dot}

The complete Hamiltonian $H$ specified above is infinite-dimensional. However, if the temperature $T$ is low and the superconducting gap $\Delta$ is large, then one may simplify the Hamiltonian by eliminating the superconducting quasiparticles from the description. Technically, this is done by integrating out the quasiparticles using second-order perturbation theory in the SC-QD tunneling term $H_{\text{T,SC}}$. This procedure yields a 16-dimensional low-energy effective Hamiltonian for the double QD, which describes the superconducting proximity effect of the SC lead on the double QD. Here, we describe this effective Hamiltonian and the procedure to obtain it.

For this, we consider the Hamiltonian without the N leads, $H_{\text{QD}}+ H_{\text{SC}} + H_{\text{T,SC}} + H_{\text{IT}}$. (We will take into account the N leads later to describe transport.) Assuming $\Delta \gg U$ and further neglecting the QD-SC tunneling $H_{\text{T,SC}}$, the 16-dimensional quasiparticle-free low-energy subspace is energetically well-separated from other states containing a finite number of quasiparticles. The low-energy subspace is spanned by the \emph{product basis}, the products of particle-number eigenstates of each QD, namely, $\left(\ket{0}_L, \ket{\uparrow}_L, \ket{\downarrow}_L, \ket{\uparrow\downarrow}_L \right) \otimes \left(\ket{0}_R, \ket{\uparrow}_R, \ket{\downarrow}_R, \ket{\uparrow\downarrow}_R \right)$, where the arrows denote the spin states of the electrons. We will use the notation $\ket{i,j} = \ket{i}_L \otimes \ket{j}_R$. We perform second-order Schrieffer-Wolff perturbation theory in the tunneling term $H_\text{T,SC}$ to obtain the effective Hamiltonian for the 16-dimensional low-energy subspace. See Supporting Information File 1. (SI) for the derivation and the validity conditions. As a result of this procedure, we find that the QD-SC tunneling $H_\text{T,SC}$ generates three coupling terms in the effective Hamiltonian: 
(i) a local (single-dot) pairing term, called \emph{local Andreev reflection} (LAR), 
(ii) a non-local (interdot) pairing term, called \emph{crossed Andreev reflection} (CAR), and 
(iii) an effective interdot tunneling term, called \emph{elastic cotunneling} (EC):
\bnen \label{eq:Heff} H_{\text{eff}} = H_{\text{QD}} + H^{\text{LAR}}_{\text{eff}} + H^{\text{CAR}}_{\text{eff}} + H^{\text{EC}}_{\text{eff}} + H_{\text{IT}}, \eden

In $H_\text{eff}$, the second and third terms read as
\bean H^{\text{LAR}}_{\text{eff}} &=& - \sum_{\alpha} \Gamma_{\text{LAR},\alpha} \left( d^{\dagger}_{\alpha\uparrow} d^{\dagger}_{\alpha\downarrow} + h.c. \right) \nonumber \\
H^{\text{CAR}}_{\text{eff}} &=& \Gamma_{\text{CAR}} \left(d^{\dagger}_{R\uparrow} d^{\dagger}_{L\downarrow} + d^{\dagger}_{L\uparrow} d^{\dagger}_{R\downarrow} + h.c. \right). \eean
The effective parameters $\Gamma_{\text{LAR},\alpha}$ and $\Gamma_{\text{CAR}}$ are related to each other on the level of the presented model, i.e. neglecting the spatial separation of the QDs. (See, e.g., Ref.~\cite{EldridgePRB2010} and SI). However in the rest of this work, we will consider $\Gamma_{\text{LAR},\alpha}$ and $\Gamma_{\text{CAR}}$ as independent parameters, since the CAR mechanism is expected to be suppressed, when a finite distance between the QDs is introduced \cite{LeijnsePRL2013}. 

The EC term $H^{\text{EC}}_{\text{eff}}$ describes single-electron tunneling between the QDs via the SC. This term, in contrast to the LAR and CAR coupling, has a strong dependence on the on-site energies $\varepsilon_L$, $\varepsilon_R$ of the QDs. For example, the EC matrix element coupling the $\ket{\uparrow,0}$ and $\ket{0,\uparrow}$ states is well approximated by
\bnen \label{eq:ecexample1} \left[ H^{\text{EC}}_{\text{eff}}\right]_{\left(0,\uparrow\right)-\left(\uparrow,0\right)} = \gamma_{\text{EC}} \left( \varepsilon_L + \varepsilon_R \right), \eden
whereas the matrix element coupling  the $\ket{\uparrow,\uparrow\downarrow}$ and $\ket{\uparrow\downarrow,\uparrow}$ states is
\bnen \label{eq:ecexample2} \left[ H^{\text{EC}}_{\text{eff}}\right]_{\left(\uparrow\downarrow,\uparrow\right)-\left(\uparrow,\uparrow\downarrow\right)} = - \gamma_{\text{EC}} \left( \varepsilon_L + \varepsilon_R + U_L + U_R \right), \eden
where the strength of the EC mechanism is characterized by the dimensionless parameter $\gamma_{\text{EC}} = \Gamma_{\text{CAR}}/\Delta$. See SI for the derivation and for the complete list of the matrix elements. 

To illustrate the dependence of these matrix elements on the QD on-site energies, we plot the two matrix elements shown in Eqs.~\eqref{eq:ecexample1} and \eqref{eq:ecexample2} in Fig.~1b, using $\Delta = 5 U$ and $\Gamma_\text{CAR} = 0.1 U$. The two matrix elements -- determined from second-order perturbation theory -- plotted in Fig.~1b vanish at $\varepsilon_L = - \varepsilon_R$ and $\varepsilon_L + U_L = - \varepsilon_R - U_R$, respectively. Of course, if such a contribution vanishes, then higher-order terms neglected here may actually be important. 
%(iii) Our perturbative treatment is reasonable as long as the irrelevant virtual states, i.e, those containing a single quasiparticle, have higher energy than the relevant, quasiparticle-free states. Throughout this work, we will focus on the on-site energy range $-1.5 U < \varepsilon_L, \varepsilon_R < 0.5 U$. Then, the above requirement translates to $\Delta > 3.5 U$. \red{ez nem csak az EC-hez kell, hanem az egesz SW-hez}

Importantly, fermion parity and spin are conserved in our effective model. This implies that the 16-dimensional effective Hamiltonian has a block structure; more precisely, there are 6 orthogonal subspaces that are not mixed by the effective Hamiltonian. These \emph{invariant subspaces} are shown in Fig.~1c. The first invariant subspace (Singlet - S, left panel of Fig.~1c) contains the five spin-singlet states with even number of electrons on the QDs: the empty and the doubly occupied states ($\ket{0,0}$,$\ket{0,\uparrow\downarrow}$,$\ket{\uparrow\downarrow,0}$,$\ket{\uparrow\downarrow,\uparrow\downarrow}$), and the spin-singlet combination of the (1,1) states, $\ket{S(1,1)}=\frac{1}{\sqrt{2}} \left(\ket{\uparrow,\downarrow}-\ket{\downarrow,\uparrow}\right)$. The second and third invariant subspaces (Doublet - D, right panel of Fig.~1c) contain the states with odd number of electrons. Since we do not account for a magnetic field, these 8 states are decomposed into two invariant subspaces with different total spin \textit{z} component $\left[\left(\ket{\uparrow,0},\ket{0,\uparrow},\ket{\uparrow,\uparrow\downarrow},\ket{\uparrow\downarrow,\uparrow}\right)\right.$ and $\left.\left(\ket{\downarrow,0},\ket{0,\downarrow},\ket{\downarrow,\uparrow\downarrow},\ket{\uparrow\downarrow,\downarrow}\right)\right]$. Each energy eigenvalue in one Doublet subspace has an equal partner in the spectrum of the other Doublet subspace. The three spin-triplet combinations of the $(1,1)$ states, i.e. $\ket{\uparrow,\uparrow}$, $\ket{\downarrow,\downarrow}$ and $\ket{T_0(1,1)} = \frac{1}{\sqrt{2}}\left(\ket{\uparrow,\downarrow}+\ket{\downarrow,\uparrow}\right)$ remain uncoupled from each other and from the other invariant subspaces, and these three states have the same energy eigenvalue. 

In Fig.~1c, the arrows visualize the tunneling-induced matrix elements coupling the basis states of the effective Hamiltonian. Note that in Fig.~1c, we use the singlet-triplet basis instead of the product basis. The tunneling processes giving rise to the coupling matrix elements indicated by red arrows on Fig.~1c, are illustrated in Fig.~1d. E.g., one of the LAR matrix elements corresponds to transferring a Cooper pair from the SC to QD$_\text{R}$ through a virtual intermediate state, in which one electron occupies QD$_\text{R}$ and one quasiparticle is present in the SC. The analogous CAR matrix element corresponds, again, to extracting a Cooper pair from the SC, but in this case the electrons end up in different QDs. Both the EC and the IT matrix elements correspond to the transfer of an electron from one QD to the other. In the case of EC, there is an intermediate virtual state with one quasi-particle in the SC, but in the case of IT, the tunneling is direct. The difference between EC and IT processes results an important difference of their matrix elements: for EC, they depend on the on-site energies [see, e.g., Eqs.~(11) and (12)], while for IT they do not. [see Eq. (8)].

In what follows, we will rely on the numerically obtained eigenvalues $E_{\chi}$ and eigenstates $\ket \chi$ of the effective Hamiltonian $H_\text{eff}$ of Eq.~\ref{eq:Heff}. Furthermore, we will use $U=U_{L}=U_{R}$ as the energy unit, $\Gamma_{\text{LAR},L}=\Gamma_{\text{LAR},R}=0.25 U$, and will focus on the parameter range $\Gamma_{\text{CAR}} \in [0,0.1] U$, $\gamma_{\text{EC}} \in [0,0.15]$, $t_{\text{LR}} \in [0,0.1] U$. To convert our results to physical units, we can use, e.g., $U=1$~meV; then the above numbers correspond to an experimentally realistic parameter set. 
%For example $\Gamma_{CAR} = 50~\mu$eV and $\Gamma_{EC} = 0.1$ would implicate a superconducting gap of $500~\mu$eV (see Eq.~11 of Supplementary Information File 1. for the definition of $\Gamma_{\text{EC}}$).

\section{Ground-state properties}

\label{sec:PD}

Here we analyze the ground-state properties of the effective Hamiltonian $H_{\text{eff}}$, namely, the ground-state degeneracy, the fermion parity and the average electron occupations of the QDs, as functions of the on-site energies $\varepsilon_L$ and $\varepsilon_R$, for different values of the non-local coupling terms CAR, EC and IT (see Fig.~2a-e). % We will focus on the limiting cases where only one of these three mechanisms are present. \red{nem igaz} 
First we analyze the fingerprints of the three non-local coupling mechanisms, one-by-one, and finally we show an example where all coupling terms are finite. We show that although CAR and EC couple different states (see Fig.~1b), they produce rather similar phase diagrams, but IT can be clearly distinguished from the previous two. In an experimental situation when one of these mechanisms is dominant, our results can be used to identify that dominant mechanism. We also note that in an actual experiment, it is challenging to effectively tune the ratio of these parameters.% independently;  %e.g., when the voltage on a barrier gate separating QD$_\text{R}$ and the SC is tuned, then that simultaneously influences (at least) the on-site energy of QD$_\text{R}$, $\Gamma_\text{LAR}$ on the right side, and all three non-local coupling parameters. \red{ennyire ne magyarazzuk mar tul... egyebkent is ha masik gate azzal az on-site energiat vissza lehet kompenzalni. en arrol beszeltem hogy a 3-4 csatolas EGYMASHOZ KEPESTI ARANYA mennyire hangolhato, pl kapu fesszel definialt QD-kat lehet messzebb tolni, igy gyengiteni a CAR-t a LAR-hoz kepest, a SC alatti palcaszakasz elektronsurusege is hangolhato valamennyire, ezzel az IT hangolhato}

\begin{figure*}
\begin{center}
\includegraphics[width=14cm,keepaspectratio]{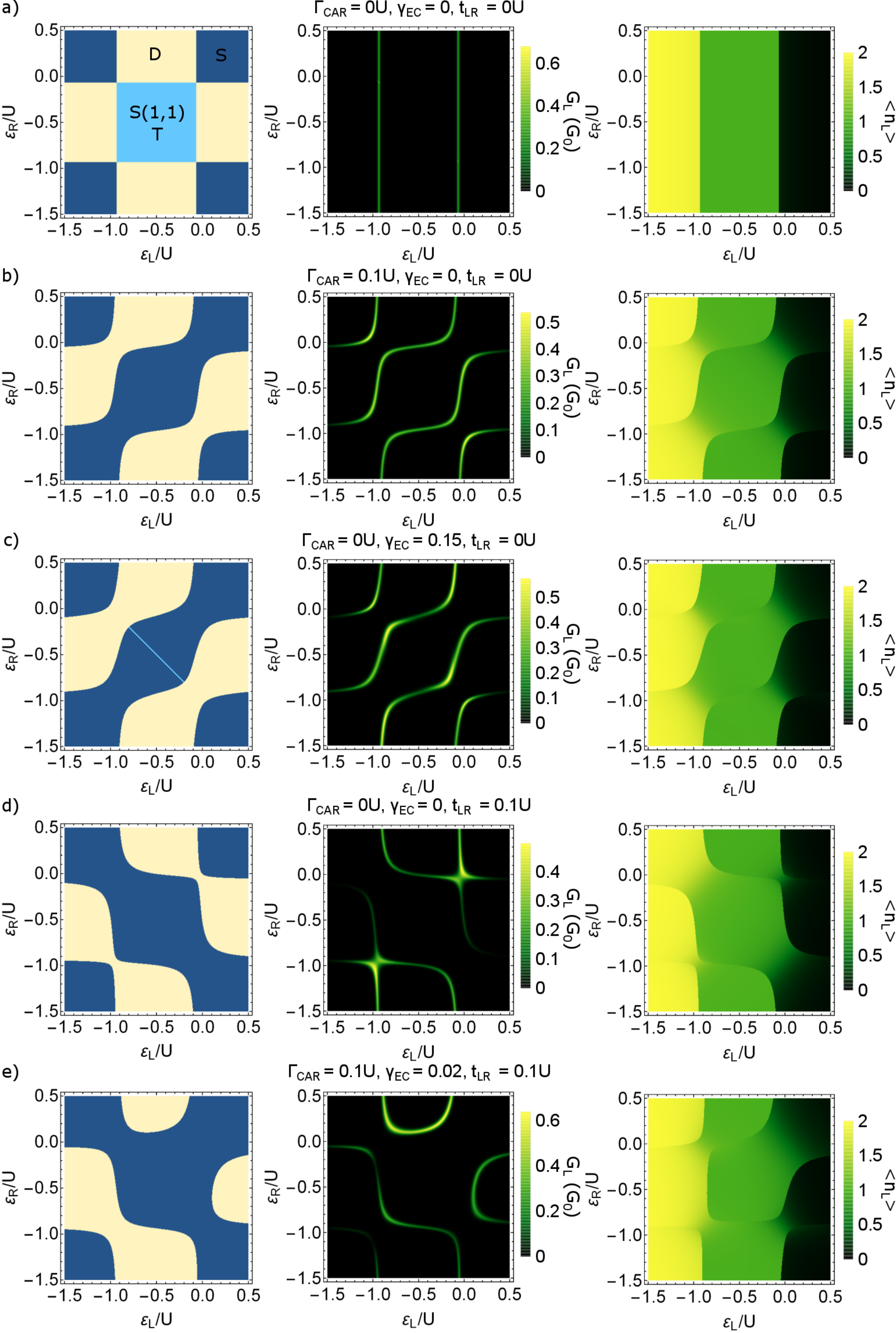}
\caption{Phase diagram and related measurable quantities of the QD-SC-QD system. Phase diagram (left column), zero-bias conductance $G_L$ of the left lead N$_\text{L}$ (middle column), and average electron occupation $\langle n_L \rangle$ of QD$_\text{L}$ (right column) are shown, for different non-local coupling configurations: a) without non-local couplings, b) only CAR, c) only EC, d) only IT, e) all three.}
\label{fig2}
\end{center}
\end{figure*}

The left column of Fig.~2 shows the phase diagram, the middle column shows the zero-bias conductance of QD$_\text{L}$, and the right column shows the average electron occupation of QD$_\text{L}$ for different $\Gamma_{\text{CAR}}$, $\gamma_{\text{EC}}$ and $t_{\text{LR}}$ values. Note that for all cases the local pairing term is finite, $\Gamma_{\text{LAR},L} = \Gamma_{\text{LAR},R} = 0.25U$. In this subsection, we discuss the phase diagram and the average electron occupation, and return to the zero-bias conductance results later, after introducing the transport model. 

The phase diagrams of the QD-SC-QD system display the dependence of two ground-state properties on the on-site energies $\varepsilon_L$ and $\varepsilon_R$: (i) the degree of degeneracy of the ground state, and (ii) the fermion parity of the ground state. For example, Fig.~2a shows the phase diagram without non-local couplings, i.e., $\Gamma_{\text{CAR}}=\gamma_{\text{EC}}=t_{\text{LR}}=0$. Different colors correspond to different ground-state degeneracies: yellow denotes a two-fold degenerate ground state, dark blue denotes a non-degenerate ground state, and light blue denotes a four-fold degenerate ground state. The ground state has even (odd) fermion parity in the dark blue and light blue (yellow) regions. Note that such a phase diagram should be regarded as the generalization of the stability diagrams of non-superconducting double QDs, see, e.g., Fig.~2 of Ref.~\cite{vanderWiel}.

In Fig.~2a, where non-local couplings are absent, the two QDs are independent, thus the phase boundaries are vertical and horizontal lines. (Recall that intradot Coulomb repulsion is neglected.) The dark blue regions correspond to a Singlet (S), unique ground state, where both QDs are in the bonding combination of the states $\ket{0}_{\alpha}$ and $\ket{\uparrow\downarrow}_{\alpha}$. Note that in the absence of non-local couplings, the state $\ket{S(1,1)}$ remains uncoupled from the other four states in the Singlet subspace, and degenerate with the three triplets $\ket{T_0(1,1)}$, $\ket{(\uparrow,\uparrow)}$, and $\ket{(\downarrow,\downarrow)}$. As a consequence, the ground-state degeneracy in the central, light blue region is four-fold. The two ground states of the yellow regions are drawn from the Doublet (D) subspace of Fig.~1b \cite{MengPRB2009}.

In general, if a non-local coupling is turned on, then the light blue regions (four-fold degenerate ground state) disappear, see Fig.~2b-e. The reason for that is as follows. Due to the non-local coupling, the state $\ket{S(1,1)}$ couples to the other four Singlet states $\left( \ket{0,0},\ket{0,\uparrow\downarrow},\ket{\uparrow\downarrow,0},\ket{\uparrow\downarrow,\uparrow\downarrow} \right)$. Therefore, the energy of the lowest-energy Singlet eigenstate will be lower than the energy of the triplets. This mechanism leaves two possibilities for the ground state: non-degenerate Singlet, or a two-fold degenerate Doublet. The only exception is a line, when EC is the only finite non-local mechanism, where the four-fold degeneracy is preserved, which we discuss below.
%the appearance of the fourfold degenerate ground state in the case when the only finite non-local mechanism is EC, which we discuss below \red{ezt se itt mondanam el}. 

The case of finite CAR coupling is presented in Fig.~2b with $\Gamma_{\text{CAR}} = 0.1 U$. Besides the disappearance of the four-fold degenerate light blue region, another difference compared to Fig.~2a is the merging of the Singlet phase regions along the the \textit{diagonal} (i.e., the $\varepsilon_L = \varepsilon_R$ line) and the merging of the Doublet phase regions parallel to the diagonal. These features are consequences of the CAR coupling, and can be understood via simple perturbative arguments. 

For example, consider the top left quadruple point in the phase diagram of Fig.~2a. In that point, the most relevant Doublet states are $\ket{\uparrow \downarrow,\sigma}$ and $\ket{\sigma,0}$. These states are coupled directly by CAR, as shown in Fig.~1b; as a consequence, the bonding combination of these will form the Doublet ground state, with an energy lowered by $\sim \Gamma_\text{CAR}$ due to the non-local coupling. On the other hand, the most relevant Singlet states are $\ket{\uparrow \downarrow,0}$ and $S(1,1)$, which are not coupled directly by CAR, see Fig.~1b. In conclusion, the Doublet ground state will have a lower energy than the Singlet ground state in the top left quadruple point, explaining the merge of the Doublet phase parallel to the diagonal in Fig.~2b. Similar considerations apply to the other three quadruple points of Fig.~2a.

Consider now the case when the only non-local coupling mechanism in the setup is EC. For this case, we can infer the ground-state character from a perturbative consideration similar to the one above. This consideration yields the expectation that the Singlet (Doublet) phases would merge along (parallel to) the \textit{skew diagonal} (i.e., the line $\varepsilon_R = U - \varepsilon_L$) of the phase diagram, in contrast to the case of CAR. However, the numerically evaluated phase diagram for this case ($\gamma_{\text{EC}} = 0.15$), plotted in Fig.~2c, shows that the phase diagram is actually very similar to Fig.~2b. 

To understand this surprising feature, we have to (i) go beyond the previous first-order perturbative analysis, and (ii) go beyond the qualitative arguments based on the selection rules of Fig.~1b, i.e., taking into account the on-site energy dependence of the EC coupling matrix elements, exemplified in Eqs.~(11) and (12). For example, consider the top right quadruple point in Fig.~2a. Here, the lowest-energy states are dominantly $\ket{0,0}$, $\ket{0,\sigma}$, $\ket{\sigma,0}$ and $\ket{\sigma,\sigma'}$, from which the Doublet states $\ket{0,\sigma}$ and $\ket{\sigma,0}$ should be coupled directly by EC (see Fig.~1b), but the corresponding matrix elements $\left[ H^{\text{EC}}_{\text{eff}}\right]_{\left(0,\sigma\right)-\left(\sigma,0\right)}$ are proportional to $\varepsilon_L + \varepsilon_R$ (see Eq.~(11)), and hence are strongly suppressed.
On the other hand, the two other states, $\ket{0,0}$ and $\ket{\sigma,\sigma'}$ are coupled by LAR and EC in second order via intermediate states, by (cf. Fig.~1b)
\bne \ket{0,0} \stackrel{\text{LAR}}{\leftrightarrow} \ket{\uparrow\downarrow,0}, \ket{0,\uparrow\downarrow} \stackrel{\text{EC}}{\leftrightarrow} \ket{S(1,1)} \ede
which includes the $\left[ H^{\text{EC}}_{\text{eff}}\right]_{\left(0,\uparrow\downarrow\right)-\left(\sigma,\bar{\sigma}\right)}$ matrix element, which is not 
suppressed. Hence, at the top right quadruple point, this second-order hybridization results in a lowered energy of the Singlet ground state. Similar considerations explain the features of the phase diagram in Fig.~2c at all four quadruple points. 

A difference between Fig.~2b and 2c is the presence of a light blue skew diagonal line at the central region of Fig.~2c. Along this line, the EC matrix element $\left[ H^{\text{EC}}_{\text{eff}}\right]_{\left(0,\uparrow\downarrow\right)-\left(\sigma,\bar{\sigma}\right)}$ vanishes (see Eq. 13 in SI), and therefore the state $\ket{S(1,1)}$ is decoupled from the other Singlet states, and remain degenerate with the triplets, preserving the four-fold degeneracy. %These four degenerate states form the ground states along the light blue line in Fig.~2c.

Fig.~2d shows the phase diagram for the case when the only non-local coupling mechanism is IT, for $t_{\text{LR}}=0.1 U$. Here, the Singlet and Doublet phases merge parallel to the skew-diagonal. This is explained by arguments analogous to the first-order perturbative considerations outlined above, keeping in mind that the non-local coupling matrix elements of IT do not depend on the QD on-site energies. 

In conclusion, if we assume that only one non-local coupling mechanism is present, then CAR and EC produce rather similar phase diagrams, but they can be clearly distinguished from the case of IT.

All phase diagrams a-d of Fig.~2  are symmetric in two ways: (i) for the transformation $(\varepsilon_L,\varepsilon_R) \mapsto (\varepsilon_R,\varepsilon_L)$, and (ii) for the transformation $(\varepsilon_L,\varepsilon_R) \mapsto (-U_L-\varepsilon_L,-U_R-\varepsilon_R)$. The property (i), which we call the \textit{left-right symmetry}, originates from the symmetric choice of local parameters, i.e., $U_L = U_R$, $\Gamma_{\text{LAR,L}} = \Gamma_{\text{LAR,R}}$. The property (ii) is the result of a \textit{particle-hole symmetry} of the system, as discussed in the following.

A particle-hole transformation converts the filled electron states to empty ones and vica versa, i.e. exchange the role of creation and annihilation operators. Eight different particle-hole transformations are introduced and discussed in SI. One example is the transformation (iv) in Table 1 of SI, corresponding to the mapping $d^{\dagger}_{L\sigma} \rightarrow \left( -1 \right)^{\sigma} d_{L\sigma}$, $d^{\dagger}_{R\sigma} \rightarrow -\left( -1 \right)^{\sigma} d_{R\sigma}$. Each transformation can be represented as a unitary transformation $W$ on the 16 dimensional Fock space. For each Hamiltonian terms $H(\varepsilon_L,\varepsilon_R)$, these transformations connect the inverted points of the phase diagram (see SI for details), namely
%Different unitary transformations $W$ of the Fock space can be defined as particle-hole transformations. \red{Nem derul ki, hogy egy uniter trafot mikor nevezel PHT-nek} Eight different particle-hole transformations are introduced and discussed in SI. One example is the transformation (iv) in Table 1 of SI, to be denoted as $W_\text{iv}$ \red{nem hasznalod kesobb}, corresponding to the mapping $d^{\dagger}_{L\sigma} \rightarrow \left( -1 \right)^{\sigma} d_{L\sigma}$, $d^{\dagger}_{R\sigma} \rightarrow -\left( -1 \right)^{\sigma} d_{R\sigma}$. For certain Hamiltonian terms $H(\varepsilon_L,\varepsilon_R)$, these transformations connect the inverted points of the phase diagram (see SI for details), namely \red{nyil jobb lenne, vagy egy prefaktor; az hogy connect meg nem implikalja, hogy egyenlo}
\begin{widetext}
 \bean
 \label{eq:phs}
 W H(\varepsilon_L,\varepsilon_R;U_L,U_R) W^\dag \propto H(-\varepsilon_L-U_L,-\varepsilon_R-U_R;U_L,U_R )
 \eean
\end{widetext}
In this sense, these transformations correspond to an inversion in the phase diagram to the central point $(\varepsilon_L,\varepsilon_R) = (-U_L/2,-U_R/2)$, usually called \textit{particle-hole symmetric point}. We say that the transformation $W$ is the particle-hole symmetry of the Hamiltonian term $H$, if Eq.~\eqref{eq:phs} is an equality. All coupling Hamiltonians $H^{\text{LAR}}_{\text{eff}}$, $H^{\text{CAR}}_{\text{eff}}$, $H^{\text{EC}}_{\text{eff}}$ and $H_{\text{IT}}$ have a few such particle-hole symmetries, but each term has a different set of those, see Table 1 of SI. If there exists a \emph{single} particle-hole transformation that is a particle-hole symmetry of \emph{all} coupling terms forming the Hamiltonian, then the phase diagram (along with other quantities) reflects the particle-hole symmetry. 
For example, the phase diagram in Fig.~2d, where IT is the only non-local coupling mechanism, shows particle-hole symmetry, since transformation (iv) in Table 1 of SI is a particle-hole symmetry of the Hamiltonian from which $H^{\text{CAR}}_{\text{eff}}$ and $H^{\text{EC}}_{\text{eff}}$ are omitted. 

Finally, consider the general case, having all non-local couplings finite, $\Gamma_{\text{CAR}} = t_{\text{LR}} = 0.1 U$ and $\gamma_{\text{EC}}=0.02$. The phase diagram for this case is shown in Fig.~2e. First, the left-right symmetry is apparent, and it is still a consequence of the left-right symmetric choice of the parameter values. Second, the particle-hole symmetry is absent in Fig.~2e. That is consistent with the fact that none of the eight particle-hole transformations is a particle-hole symmetry of all terms in this general Hamiltonian (see Table 1 in SI). 

%The previous argument of the energy lowering of the $\ket{S(1,1)}$ compared to the triplets, resulting in two different character of ground states is also valid here.  The different direction of the merging of phase regions close to $\varepsilon_L = \varepsilon_R = 0$ and $\varepsilon_L = \varepsilon_R = -1$ is the result of the interference of EC and IT. While in one case the non-suppressed matrix element of EC adds to IT, for the other case it is subtracted. 

The phase boundaries between ground states of different fermion parities can be mapped by low-energy transport \cite{vanderWiel}. 
%This method is analogous to the mapping of the triple points of the honeycomb stability diagram of a normal double QD.
The expected zero-bias conductance, with peaks following the even-odd phase boundaries, is illustrated by the middle column of Fig.~2, and will be discussed in more detail below. A drawback of transport analysis is that the coupling to the electrodes lead to the broadening of conductance peaks (which effect is neglected from the model presented here). One may reduce broadening by decreasing the coupling at the price of decreasing the currents too. 

Charge sensing \cite{Sprinzak,ElzermanPRB2003} is another method to map out the boundaries of the phase diagram, as we illustrate in the right column of Fig.~2. A charge sensor is usually engineered to be mostly sensitive to the average electron occupation of one of the quantum dots, say, QD$_L$. Compared to the conductance measurement through the QD-SC-QD system, charge sensing has the advantage of conceptual simplicity, and the measurability without additional N leads attached to the QD-SC-QD system; but might have the disadvantage of a more complex device design, since the charge sensor is an additional device element. Similar methods, yielding information related to average electron occupation, are based on reflectometry with electromagnetic radiofrequency signals \cite{SillanpaaPRl2005,DutyPRL2005} or microwave resonators\cite{FreyPRL2012,PeterssonNature2012}; those are not discussed further in this work. 

The ground-state average electron occupation in QD$_L$ is expressed as $\left\langle n_L \right\rangle = \left\langle \sum_{\sigma} d^{\dagger}_{L\sigma} d_{L\sigma} \right\rangle$. We plot $\langle n_L \rangle$ as the function of the on-site energies $\varepsilon_L$ and $\varepsilon_R$ in the right column of Fig.~2, for the parameter values providing the previously discussed phase diagrams.

In the absence of non-local couplings (Fig.~2a), the QDs are independent, therefore $\left\langle n_L \right\rangle$ does not depend on $\varepsilon_R$. We emphasize, since it is not apparent in the $\langle n_L \rangle$ density plot in Fig.~2a,  that values of $\langle n_L \rangle$ are \emph{not} restricted to the integer values 0, 1 and 2: this is because the LAR mechanism provides coherent coupling within a given fermion-parity sector between states with different electron numbers, see Fig.~1b. In fact, $\langle n_L \rangle$ as a function of $\varepsilon_L$ slightly decreases in the yellow $\left(\langle n_L \rangle \approx 2\right)$ and black $\left(\langle n_L \rangle \approx 0\right)$ regions in the right panel of Fig.~2a, its value is strictly $\langle n_L \rangle = 1$ in the green region, and jumps abruptly at the boundaries. 

The non-local couplings are switched on in Figs.~2b-e. Similarly to Fig.~2a, the average electron occupation $\langle n_L \rangle$ decreases as $\varepsilon_L$ is increased, and the jump locations in $\left\langle n_L \right\rangle$ follow the phase boundaries. The jumps are more pronounced along the vertical phase boundaries, i.e. in a charge sensing experiment the measurement of QD$_{\text{L}}$ maps out the vertical phase boundary lines more efficiently. Due to the finite non-local couplings, the variation of the average electron occupation within a given fermion-parity sector is smooth, as shown in Fig.~2b-e.

Due to the left-right symmetry, for all cases presented here, the average electron occupation of QD$_{\text{R}}$, that is, $\left\langle n_R \right\rangle$, can be obtained by mirroring $\left\langle n_L \right\rangle$ to the diagonal. Therefore, in the $\left\langle n_R \right\rangle$ map the horizontal phase boundary lines are more pronounced. This allows for the measurement of the phase boundaries by measuring the occupation of the two QDs independently. 

In this subsection, we have shown that on contrary to the naive expectations, the CAR and the EC mechanisms produce rather similar phase diagrams as the function of $\varepsilon_L$ and $\varepsilon_R$, but IT can be clearly distinguished from the previous two mechanisms. Furthermore, the measurement of the average electron occupation of the QDs allows for determining the phase boundaries, even in the absence of normal electrodes tunnel-coupled to the QD-SC-QD system. 

\section{Transport calculation}

As pointed out earlier, a charge-sensing measurement is demanding, since the addition of the charge sensor complicates device fabrication. However, the ground-state phase diagrams discussed above can also be explored experimentally by electronic transport measurements, utilizing two additional N leads besides the SC lead (see Fig.~1a) and low bias voltages. We will demonstrate this using the results shown in the middle column of Fig.~2. In addition, transport measurements using a sufficiently large bias voltage allow to determine energy gaps above the ground state. We will show (see Fig.~4) that such finite-bias measurements can distinguish the CAR-dominated and EC-dominated cases, i.e., the two cases that are not distinguished by the ground-state properties shown in Figs.~2b,c. 

The transport setup we wish to describe is shown in Fig.~1a. Throughout this work, we assume that the SC lead is grounded, $\mu_\text{SC} = 0$, and the two N leads are biased symmetrically, $\mu_\text{NL} = \mu_\text{NR} = \mu_\text{N}$, with the convention that for positive (negative) $\mu_\text{N}$, the electrons tend to flow into (out from) the SC lead.

Our transport model is based on the effective Hamiltonian $H_{\text{eff}}$ of Eq.~\eqref{eq:Heff} describing the QD-SC-QD system. In addition, here we also take into account the N-lead Hamiltonians and the lead-QD tunneling Hamiltonians, that is, $H_{\text{N}} + H_{\text{T,N}}$. We describe the electronic transport in this device using a classical master equation, where the tunnel rates between the N leads and the QD-SC-QD system are obtained perturbatively, from Fermi's golden rule. 

The classical master equation describes the time evolution of the occupation probabilities $P_\chi(t)$ of the 16 energy eigenstates $\ket{\chi}$ of $H_\text{eff}$, and reads
\bnen \label{eq:master} \frac{dP_{\chi}}{dt} = \sum_{\chi' \neq \chi} \left( W_{\chi  \chi'} P_{\chi'} - W_{\chi' \chi} P_{\chi} \right), \eden
with the normalization condition $\sum_{\chi} P_{\chi} = 1$. The transition rates $W_{\chi \chi'} = \sum_{\alpha\sigma} \left( W_{\chi \chi'} \left( d^{\dagger}_{\alpha\sigma} \right) + W_{\chi \chi'} \left( d_{\alpha\sigma} \right) \right)$ are obtained from the leading-order term in Fermi's golden rule as 
\bean W_{\chi \chi'} \left( d^{\dagger}_{\alpha\sigma} \right) &=& \Gamma_{N\alpha} \abs{\bra{\chi} d^{\dagger}_{\alpha\sigma} \ket{\chi'}}^2 f \left( E_{\chi} - E_{\chi'} - \mu_{N\alpha} \right) \nonumber, 
\\ W_{\chi \chi'} \left( d_{\alpha\sigma} \right) &=& \Gamma_{N\alpha} \abs{\bra{\chi} d_{\alpha\sigma} \ket{\chi'}}^2 f \left( E_{\chi} - E_{\chi'} + \mu_{N\alpha} \right), 
\eean
where $\alpha \in \{L,R\}$ is the lead index, $\Gamma_{N\alpha} = \pi \rho_{N\alpha} t^2_{N\alpha}$ is a characteristic tunneling rate between the normal lead N$_{\alpha}$ and the SC-QD-SC system, $\rho_{N\alpha}$ is the density of states at the Fermi-energy in the lead N$_{\alpha}$, and $f \left( x \right) = 1/\left( e^{x/k_B T} + 1 \right)$ is the Fermi-function.

We calculate the stationary ($dP_\chi / dt = 0$) solution of Eq.~\eqref{eq:master} to obtain the steady-state occupation probabilities $P_\chi$, and use the latter to evaluate the steady-state current in lead $N_\alpha$ via
\bnen I_{\alpha} = \frac{e}{\hbar} \sum_{\chi'\sigma} \left( W_{\chi' \chi} \left( d_{\alpha\sigma} \right) - W_{\chi' \chi} \left( d^{\dagger}_{\alpha\sigma} \right) \right) P_{\chi}. \eden
Since we use a leading-order Fermi's golden rule, each non-zero transition rate corresponds to the tunneling of a single electron between a lead and the QD-SC-QD system, therefore connects states with different fermion parities. The differential conductance of lead N$_{\alpha}$ is calculated by numerically differentiating the current by the chemical potential of the normal leads, i.e. $G_{\alpha} = e \frac{dI_{\alpha}}{d\mu_\text{N}}$. %This approach is reliable as long as the chemical potential shift used for the numerical differentiation is much smaller than the thermal energy scale, $d \mu_N \ll k_\text{B} T$. \red{Ebbol profital az olvaso? egyebkent se igaz }
We plot the conductance in the units of the conductance quantum $G_0=2e^2/h.$

In the following, we will use $\Gamma_{NL} = \Gamma_{NR} = \Gamma_N = 0.005 U$ and $k_{\text{B}} T = 0.005 U$. If our energy unit is $U = 1$~meV, then $\Gamma_{N} = 5~\mu$eV. Note that in our model, the currents and differential conductances are simple linear functions of $\Gamma_{N}$, since we use a leading-order Fermi's golden rule. Furthermore, the broadening of the conductance resonances in our results is caused only by the finite temperature of $k_{\text{B}} T=0.005 U \approx k_{\text{B}} \cdot 60$~mK, since the presented model neglects the life-time broadening. Experimentally, life-time broadening of the conductance resonances might be dominant over the thermal broadening; in that case, the linewidth can be reduced by decreasing the tunnel coupling to the N leads, at the price of suppressing the currents.  

\section{Results of the transport simulation}

Here, we present the results we obtained from the transport model of the previous section. First, we discuss how to experimentally map the phase boundaries using zero-bias conductance measurements, as presented in Fig.~2. Second, we demonstrate that finite-bias differential conductance measurements provide a means to distinguish a system dominated by CAR from one dominated by EC. Third, we demonstrate and analyze the appearance of negative differential conductance in our setup, which is often attributed to the triplet blockade \cite{EldridgePRB2010,TrochaPRB2015} and the CAR mechanism; here we show that not only CAR but any of the three non-local coupling mechanism can result in triplet blockade and a corresponding negative differential conductance. 

\subsection{Zero-bias conductance}

\label{sec:ZBC}

The zero-bias conductance of lead N$_{\text{L}}$, $G_L$ is shown in the middle column of Fig.~2, for the five previously discussed cases, i.e. in the absence of non-local couplings (a), having only one of them turned on (b,c,d), and in a general case of having all three finite (e). For all five cases, $G_L$ shows a resonant enhancement along the phase boundaries. The conductance in Fig.~2b-e shows further enhancement in the vicinities of the quadruple points seen in the left panel of Fig.~2a.

Due to the left-right symmetric choice of the parameter values, similarly to the average electron occupation, the conductance $G_R$ of QD$_{\text{R}}$ can be obtained from $G_L$ by mirroring the latter to the diagonal $\varepsilon_L = \varepsilon_R$ line.

In conclusion, the zero-bias measurement is a sufficient tool to locate the phase boundaries in the $\varepsilon_L - \varepsilon_R$ plane, and to determine whether the interaction between the dots is due to the proximity of the superconductor (CAR or EC dominates over IT), or not (IT dominated).

\subsection{Finite-bias conductance}

Here, first we show how the presence non-local couplings affect the finite-bias transport for a general case, then we show that CAR and EC provide different fingerprints, and therefore these measurements can distinguish between a CAR-dominated setup and an EC-dominated setup. 

Fig.~3a shows the finite-bias conductances $G_L$ and $G_R$ of the two leads, in the absence of non-local couplings, for a fixed value $\varepsilon_R = -1.2U$, as the function of $\varepsilon_L$ and the bias voltage $\mu_N$. Fig.~3b shows the same quantities, for a general case when all three non-local couplings are switched on. 

\begin{figure*}
\begin{center}
\includegraphics[width=12.5cm,keepaspectratio]{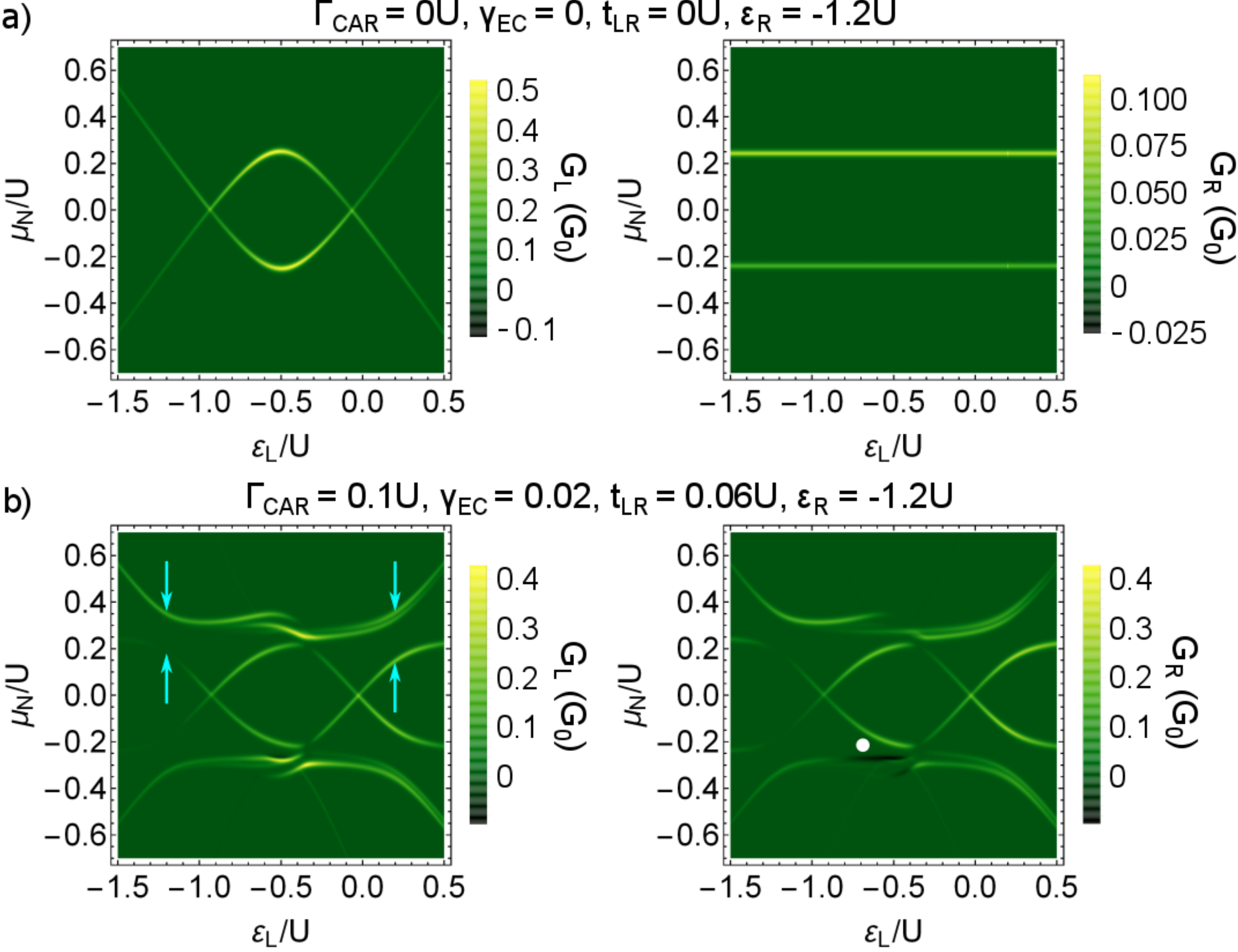}
\caption{
Hybridization of the two quantum dots of the Cooper-pair splitter due to the non-local coupling mechanisms. Differential conductances $G_L$ and $G_R$ of the Cooper-pair splitter are shown in the left and right panels, respectively, for $U_L = U_R = U$, $\Gamma_{LAR,L}=\Gamma_{LAR,R}=0.25 U$, and fixed $\varepsilon_R = -1.2 U$, as a function of $\varepsilon_L$ and bias voltage $\mu_N$. 
a) Without non-local couplings. Differential conductance shows the single-dot Andreev bound states formed on the dots. 
b) With non-local couplings. Their presence leads to the hybridization of the Andreev bound states on the dots, indicated by the appearance of anticrossings marked with arrows. Parameters: $\Gamma_{\text{CAR}}=0.1 U$, $\gamma_{\text{EC}}=0.02$, $t_{\text{LR}}=0.1 U$. For better visibility, different color scales were used for different panels.}
\label{fig3}
\end{center}
\end{figure*}

Without non-local couplings (Fig.~3a), Andreev bound states (ABSs) are formed on each QD due to the coupling to the superconductor. The ABSs on the two QDs are independent. In Fig.~3a, the conductance $G_L$ in the left panel shows an eye-shaped resonance with two crossing points at zero bias. This characteristic resonance is the usual fingerprint of an ABS in a single QD (see e.g. \cite{MengPRB2009}). The resonance maps the excitation energy between the two lowest-energy eigenstates of the QD$_L$-SC subsystem: the bonding linear combination of the singlet-like state, $\ket{-}_L = \alpha_- \ket{0}_L + \beta_- \ket{\uparrow\downarrow}_L$, and the doublets, $\ket{\sigma}_L$. In the central region, the doublets are the ground states, while on the sides, the ground state is the singlet $\ket{-}_L$. Note that the higher-energy antibonding singlet state, $\ket{+}_L =\alpha_+ \ket{0}_L + \beta_+ \ket{\uparrow\downarrow}_L$, can be accessed from the doublet states, but the corresponding conductance resonance lies out of the energy window used here. Note also that in our model, transitions between the two singlet states are forbidden, since these two states both have even fermion parity. 

As noted earlier, the positive $\mu_N$ sides of the finite-bias conductance plots of Fig.~3 correspond to transport by adding electrons from the N$_{\text{L}}$ to QD$_{\text{L}}$, while the negative $\mu_N$ sides represent the opposite processes. Although the finite-bias conductance plots are not symmetric to the zero-bias axis, the conductance lines are positioned symmetrically, since (i) the spectrum does not depend on $\mu_N$ and (ii) transition between the eigenstates with different fermion parity is possible either by adding or by removing one electron. Generally, the tunnel rate for adding an electron is different from the tunnel rate for removing an electron, therefore the heights of the conductance resonances at positive and negative bias are different.

A complete transport cycle constitutes of the transport of two electrons, e. g., for positive bias, the first electron enters the QD with positive energy, bringing it to an excited state, while the second electron enters with negative energy, making the system relax back to its ground state. These two electrons, which reside in the leads at the beginning of the cycle, are absorbed by the SC as a Cooper pair by the end of the cycle. Analogously, for negative bias, the first (second) electron leaves the QD with negative (positive) energy.

The finite-bias conductance plots in the absence of non-local couplings (Fig.~3a) have an inversion symmetry, $G_L \left( \varepsilon_L, \mu_N \right) = G_L \left( - U_L - \varepsilon_L, - \mu_N \right)$. We attribute this to the particle-hole symmetry discussed above, with an extension of the particle-hole transformations to the lead electrons. In the presence of non-local couplings, if the Hamiltonian is particle-hole symmetric, then the inversion symmetry $G_L \left( \varepsilon_L, \varepsilon_R, \mu_N \right) = G_L \left( - U_L - \varepsilon_L, -U_R - \varepsilon_R, - \mu_N \right)$ would be reflected in the finite-bias conductance plots (not shown).

In the finite-bias conductance plot of QD$_{\text{R}}$ only two horizontal lines are present (see the right panel in Fig.~3a). This is due to the independence of the QDs: $\varepsilon_L$ has no effect on the transport via QD$_{\text{R}}$. The position and the amplitude of the lines is the same as the ones in the left panel of Fig.~3a, at $\varepsilon_L = -1.2 U$.

When the non-local couplings are switched on, they induce a crosstalk between the two QDs, resulting in a strong modification of the finite-bias conductance plots. This is shown in Fig.~3b, for a general case with all three non-local coupling being finite, $\Gamma_{\text{CAR}} = 0.1 U$, $\gamma_{\text{EC}} = 0.02$ and $t_{\text{LR}} = 0.06 U$. The previously discussed features, i.e., the eye-shaped resonances in $G_L$ and the horizontal lines in $G_R$, appear in \emph{both} finite-bias conductance plots of Fig.~3b. In addition, anticrossings between conductance resonances are openened by the non-local couplings wherever the excitation energies of the two uncoupled dots would coincide.  The two most pronounced anticrossings are marked with arrows in the left panel of Fig.~3b, open at $\varepsilon_L = \varepsilon_R = -1.2 U$ and $\varepsilon_L = - U - \varepsilon_R = 0.2 U$.

The analysis of the conductance resonances as functions of the on-site energies allows one to distinguish between the coupling terms. This is shown in Fig.~4, where the finite-bias conductance along the skew diagonal $\varepsilon_R = - U - \varepsilon_L$ of the phase diagram is shown, for only one finite non-local coupling. Note that along this skew diagonal, we have $G_L = G_R$. Panel a), b) and c) shows the case of CAR, EC and IT coupling, with $\Gamma_{\text{CAR}}=0.1 U$, $\gamma_{\text{EC}}=0.15$ and $t_{\text{LR}}=0.1 U$, respectively. The conductance resonances in Fig.~4a and Fig.~4b show an appreciable splitting, which is absent in the right panel. The size of the splitting is constant for CAR, while for EC it is the largest at the particle-hole symmetric point, $\varepsilon_L = \varepsilon_R = - U/2$ and decreases further from it. This difference between the finite-bias conductance can used to identify whether the dominant non-local coupling mechanism is CAR or EC. As expected from the qualitatively different nature of the phase diagrams of Fig.~2 (b and c versus d), the case of IT is also qualitatively different in terms of finite-bias conductance (Fig.~4 a and b versus c).

%In the case of CAR (Fig.~and EC, the anticrossing constitutes of three lines. The lower branch and one of the higher branch corresponds to the excitation from the ground state to the first (second) excited state, while the third line corresponds to the excitation from the first excited state to the third one.
%\blue{nekem nem vilagos hogy milyen branch-ekrol van szo;
%megmutathatnank esetleg a relevans alacsonyenergias allapotok
%spektrumat}

\begin{figure*}
\begin{center}
\includegraphics[width=17cm,keepaspectratio]{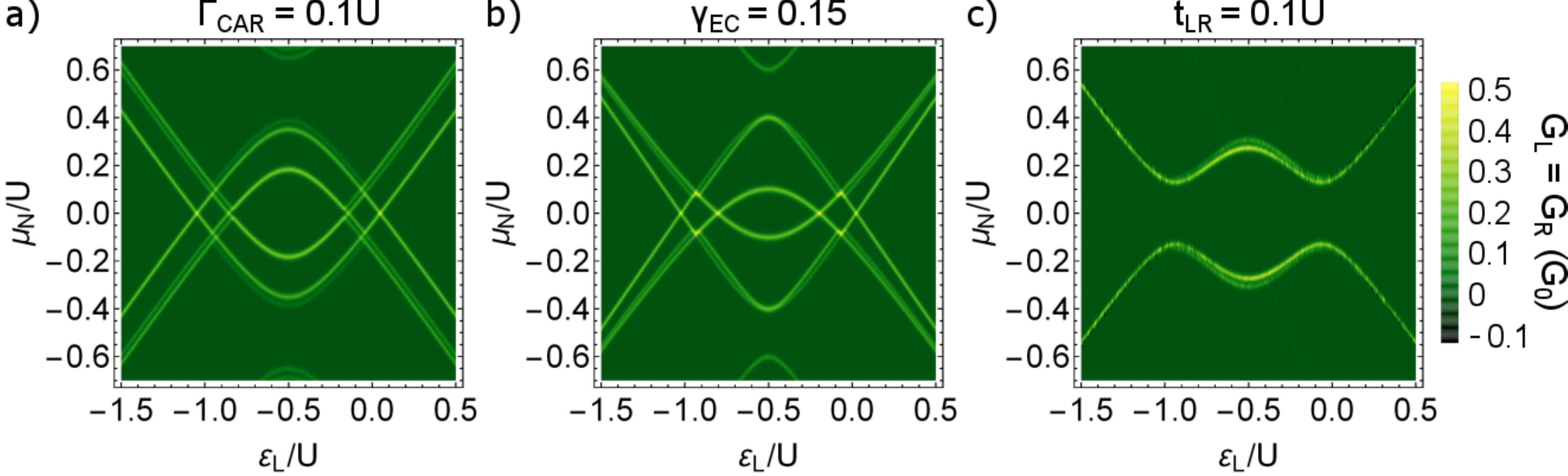}
\caption{Differential conductance as a fingerprint of the non-local coupling mechanism. Differential conductance of the QD-SC-QD system is shown, for the previously used parameters $U_L=U_R=U$ and $\Gamma_{LAR,L}=\Gamma_{LAR,R}=0.25 U$, along the skew-diagonal line $\varepsilon_R = -U - \varepsilon_L$. Only one non-local coupling mechanism is non-zero for each panel.  
a) $\Gamma_{\text{CAR}}=0.1 U$, 
b) $\gamma_{\text{EC}}=0.15$
c) $t_{\text{LR}}=0.1 U$. 
The dependence of the splitting on $\varepsilon_L$ is qualitatively different in a) and b), implying that experimental conductance data can be used to determine whether CAR or EC is the dominant non-local coupling mechanism. }
\label{fig4}
\end{center}
\end{figure*}

Further finite bias data and discussion on the features of Figs.~3\&4 is provided in SI.

\subsection{Triplet Blockade}

In certain cases, the finite-bias conductance plots show line-shaped regions of negative differential conductance (NDC). Examples, marked with white circles, are shown in the right panel of Fig.~3b and in Fig.~5. Further examples are shown on Fig.~S1\&S2 in SI.

The NDC lines appear when a so-called blocking state becomes energetically available as $\mu_N$ reaches its excitation energy. The blocking state starts to be populated, but the rates of transitions out of this state are small, therefore the population is accumulating in the blocking state, reducing the current. When the blocking states are spin-triplets, then this effect is called the triplet blockade. The triplet blockade is often referred to as a hallmark of the CAR coupling in the literature: it is argued that the spin incompatibility of the triplet state in the QDs and the spin-singlet Cooper pairs in the SC prohibit CAR coupling of the QDs \cite{EldridgePRB2010,TrochaPRB2015}. Here, we show that the presence of NDC lines is not exclusive for the CAR, but they can appear for all three non-local couplings.

Fig.~5a shows an example of triplet-blockade-related NDC lines in the CAR coupled case, for $\Gamma_{\text{CAR}}=0.1 U$ and $\varepsilon_R=- U/2$. Let us focus on the $\varepsilon_L = 0.1 U$ cut, marked with the dashed orange vertical line. For these parameter values, we show the relevant energy levels in the left panel of Fig.~5b. The ground state is a Doublet, $\ket{D} = 0.096 \ket{\sigma,0} \pm 0.959 \ket{0,\sigma} + 0.196 \ket{\uparrow\downarrow,\sigma} \pm 0.179 \ket{\sigma,\uparrow\downarrow}$, and the first excited state is a Singlet, $\ket{S} = 0.632 \ket{S(1,1)} - 0.4 \ket{0,0} - 0.193 \ket{0,\uparrow\downarrow} -0.054 \ket{\uparrow\downarrow,0} + 0.039 \ket{\uparrow\downarrow,\uparrow\downarrow}$, with a dominant $\ket{S(1,1)}$ contribution. The second excited level is the threefold degenerate triplet level. All excited states are reachable from the ground state via electron tunneling, but tunneling transitions between two different excited states are forbidden, as both states have even fermion parity. 

\begin{figure*}
\begin{center}
\includegraphics[width=17cm,keepaspectratio]{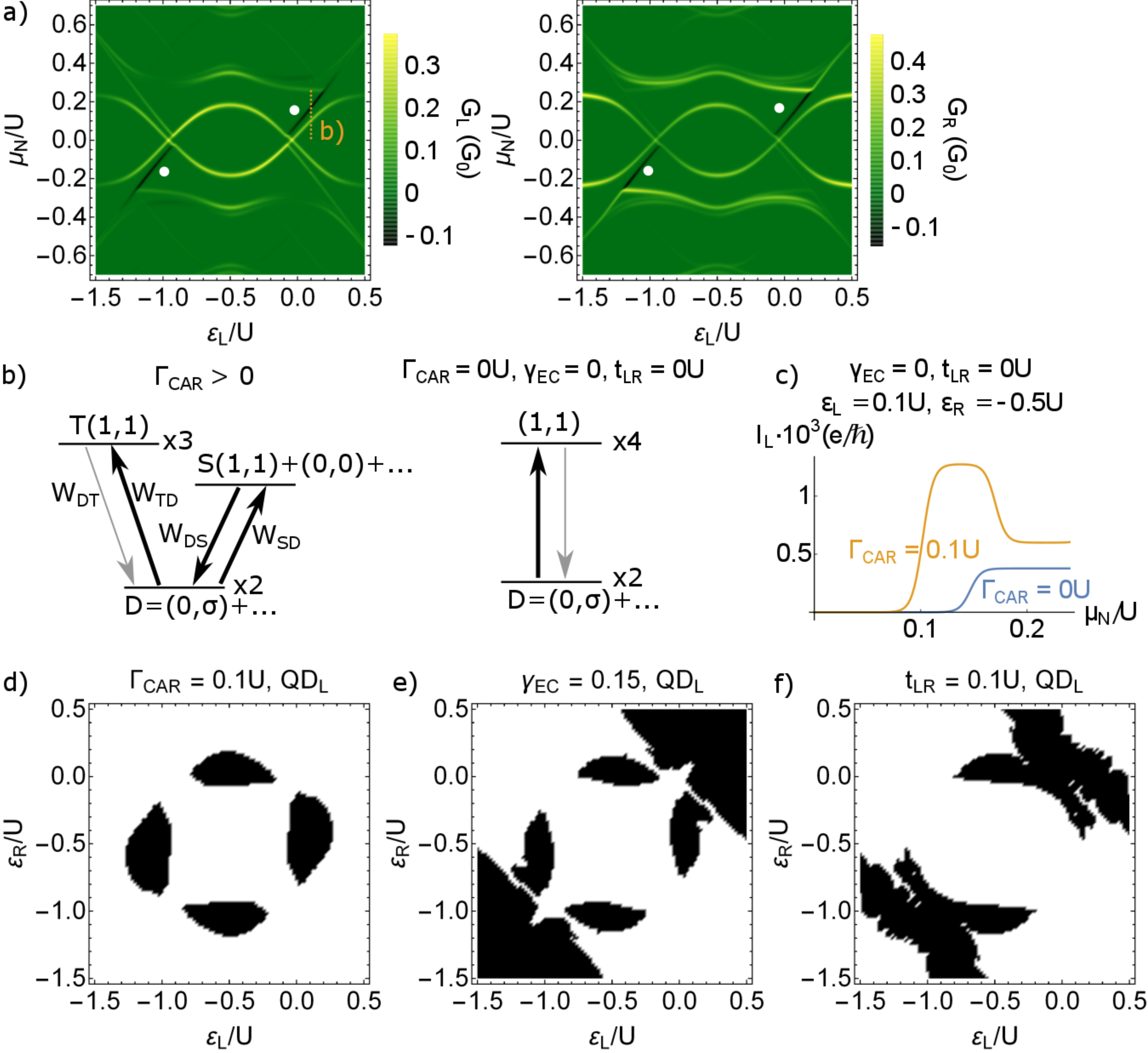}
\caption{Triplet blockade and negative differential conductance in the Cooper-pair splitter. 
a) Finite-bias differential conductances $G_L$ and $G_R$ for $\Gamma_{\text{CAR}} = 0.1 U$, $\gamma_{\text{EC}} = 0$, $t_{\text{LR}} = 0$, at $\varepsilon_R = - U/2$. 
b) Levels and transitions taking part in triplet blockade.
c) Current-voltage dependence in lead $N_L$ with (orange) and without (blue) CAR coupling, along the dotted vertical line in panel a). Introducing a non-local coupling mixes the states, hence a well conducting channel is opened. 
d-f) On-site-energy regions where triplet-blockade-induced negative differential conductance is present (black)
d) $\Gamma_{\text{CAR}}=0.1 U$,
e) $\gamma_{\text{EC}}=0.15$,
f) $t_{\text{LR}} = 0.1 U$.}
\label{fig5}
\end{center}
\end{figure*}

In the left panel of Fig.~5b, we illustrate the tunneling transitions (arrows) relevant for electron transport in this parameter point. If the bias voltage is sufficiently large to induce a transition to the Singlet excited state, then a well-conducting transport channel opens up, characterized by the uphill rate $W_{SD}$ and the downhill rate $W_{DS}$. In this transport cycle, the uphill transition, $\ket{D}\rightarrow\ket{S}$, is dominated by the $\ket{0,\sigma} \stackrel{d^{\dagger}_{L\bar{\sigma}}}{\rightarrow} \ket{S(1,1)}$ process, while the downhill transition, $\ket{S}\rightarrow\ket{D}$, is dominated by $\ket{0,0} \stackrel{d^{\dagger}_{R\sigma}}{\rightarrow} \ket{0,\sigma}$ and $\ket{\sigma,\bar{\sigma}} \stackrel{d^{\dagger}_{R\sigma}}{\rightarrow} \ket{\sigma,\uparrow\downarrow}$ processes. Here, $W_{SD}=0.383\, \Gamma_N $ and $W_{DS}=0.297\, \Gamma_N$ are the relevant transition rates. 

When the bias voltage is further increased, then the triplet states are also populated, see Fig.~5b. In this case, the  uphill transition from the ground state to the triplet is dominated by the $\ket{0,\sigma} \stackrel{d^{\dagger}_{L\sigma/\bar{\sigma}}}{\rightarrow} \ket{T(1,1)}$ process, with the rate $W_{TD}=0.465\, \Gamma_N$ rate. The downhill transitions from the triplets are possible, and their rates are dominated by the $\ket{T(1,1)} \stackrel{d^{\dagger}_{L\sigma/\bar{\sigma}}}{\rightarrow} \ket{\uparrow\downarrow,\sigma}$ and $\ket{T(1,1)} \stackrel{d^{\dagger}_{R\sigma/\bar{\sigma}}}{\rightarrow} \ket{\sigma,\uparrow\downarrow}$ processes, but these rates are small, $W_{DT}=0.019\, \Gamma_N$, more than an order of magnitude smaller than the other transition rates. This small downhill transition rate results in the accumulation of the population in the triplet states, and the reduction the Singlet and Doublet populations that would provide efficient conduction. Therefore, the net effect of the triplet states becoming available upon increasing the bias voltage is the reduction of the current, and as a consequence, the appearance of the NDC lines in the differential conductivity. 

We show the level diagram and the transitions, in the absence of CAR coupling, in the right panel of Fig.~5b. In this case, due to the absence of CAR, the $\ket{\sigma,\uparrow\downarrow}$ component is missing from the ground states, and the $\ket{0,0}$ is missing from the excited states, which are parts of the two main downhill processes enabling the $\ket{S}\rightarrow\ket{D}$ transition, as we have seen above. Therefore all four, now degenerate, $(1,1)$ states block the transport, as illustrated on the right panel of Fig.~5b.

The comparison of the two cases is shown in Fig.~5c, where the current $I_L$ in the left lead is plotted as the function the bias voltage $\mu_{N}$ with blue (orange) line without non-local coupling (with $\Gamma_{\text{CAR}}=0.1 U$) for the same $\varepsilon_L=0.1 U$ and $\varepsilon_R=- U/2$ as above. In the presence of finite CAR, as the bias voltage is increased from zero, first a well conducting channel opens around $\mu_N \approx 0.1 \, U$, and the current jumps down at $\mu_N \approx 0.17 \, U$ due to the triplet blockade. In contrast, in the absence of non-local couplings, above $\mu_N \approx 0.15 \, U$ the current is comparably small as in the blockaded case.  Note that in the presence of CAR coupling, the triplet excitation is shifted to somewhat higher energy, since the energy of the ground state is lowered due to the non-local coupling.

%As we have seen the explanation of the blockade did not involve any of the non-local terms. Indeed in the absence of non-local couplings the blocking effect is still present, which is illustrated on Fig.~5b, where the current in the left lead, $I_L$ is plotted as the function $\mu_{N}$ with blue (yellow) line without non-local coupling (with $\Gamma_{\text{CAR}}=0.1$) for the same $\varepsilon_L=0.1$ and $\varepsilon_R=-0.5$ as above. Above the excitation energy of the triplets the currents are comparably small in both cases, $I_L = 0.4-0.6 \cdot 10^{-3}$ (see $\mu_N>0.2$), showing that the blocking is present in the absence of non-local coupling as well. The difference in the presence of CAR is that a well conducting channel opens below the triplet excitation energy ($0.1<\mu_N<0.15$) via the first excited state, the singlet, which is split from the triplets. When the current drops to the blockade value a NDC line appear in the stability diagram. Note that in the presence of CAR coupling the triplet excitation is shifted to somewhat higher energy, since the energy of the ground state is lowered due to the non-local coupling.

The orange line in Fig.~5c indicates a negative differential conductance. However,  the current does not drop to zero but forms a finite plateau for $\mu > 0.17 \, U$, i.e., the triplet blockade is `incomplete', and there is a finite `leakage current'. As discussed by Trocha and Weymann in Ref.~\cite{TrochaPRB2015}, this leakage current is due to the fact that double occupancy of the QDs is allowed in this model. In contrast, in models neglecting double occupancy, e.g., by assuming infinite on-site Coulomb repulsion $U\to \infty$, this leakage current vanishes exactly \cite{EldridgePRB2010}.

As we have seen, the presence of the high current due to the well-conducting Singlet-Doublet channel requires that (i) the Singlet energy is brought below the energy of the triplets, and (ii) the $\ket{S(1,1)}$ state is mixed with the empty or double occupied states. Importantly, these conditions can be induced not only by CAR, but by any of the three non-local couplings described here. Therefore, for all the three cases, triplet-blockade-induced NDC lines can appear. In Fig.~5c-e we map the on-site energy regions where the triplet blockade is present for the three different couplings. We have marked with black those $\left(\varepsilon_L,\varepsilon_R\right)$ values where there exists a bias voltage, at which the absolute value of the current $I_L$ decreases, and the triplet occupation of the QD-SC-QD system increases by at least 0.25 simultaneously. The black regions of Fig.~5d-f are not symmetric for mirroring to the diagonal $\varepsilon_L = \varepsilon_R$ line, therefore the NDC lines are present in somewhat different $\left(\varepsilon_L,\varepsilon_R\right)$ regions for QD$_{\text{L}}$ and QD$_{\text{R}}$. Note that due to the left-right symmetry the NDC map calculated from $I_R$ would be the same as the presented ones mirrored to the diagonal.

In our setup, NDC can also appear in cases where it is not caused by triplet blockade. In the main text, we have been focusing on triplet-blockade-induced NDC, but an example for non-triplet-induced NDC is show in SI in Fig.~S1b.

In conclusion, we have shown that the triplet blockade is not exclusive for CAR, but can appear also in the presence of any of the three non-local coupling mechanisms. In fact, a blockade can arise even in the absence of non-local couplings, even though the Singlet state is also blocking in that case. The presence of a non-local coupling makes the Singlet state well conducting, and makes the triplet blockade effect easily observable in a transport experiment by the appearance of NDC lines.

\section{Summary}

We have analyzed the spectrum of a QD-SC-QD system in the presence of different non-local coupling mechanisms: CAR, EC and IT. Our aim was to calculate the effects of these on measurable quantities (phase diagrams, average electron occupations, zero-bias and finite-bias conductance), with the goal of identifying features that are characteristic for each non-local coupling mechanism. 

The phase diagram of the system can be mapped via charge-sensing or zero-bias conductance measurements. We find that that CAR and EC produces very similar phase diagrams, thus measuring the phase diagram alone would not allow to distinguish between these two non-local coupling mechanisms. However, IT produces a qualitatively different phase diagram, and hence it should be straightforward to identify if the non-local couplings are dominated by IT in a device. Furthermore, we have demonstrated that finite-bias measurements could be used to distinguish between the cases when CAR or EC dominates the non-local couplings. 

In the literature, triplet blockade is often linked to the presence of the CAR mechanism. Here we have shown that the suppression of the current due to the population of the triplet states is not specific to CAR coupling. In fact, such a current suppression can appear even without any non-local mechanism present in the device. In the presence of non-local processes, the current suppression can be interpreted as a triplet blockade, and it is observed via NDC lines, and this effect is not unique for CAR, but EC and IT can also generate it. 

We expect that the results presented here will facilitate the accurate characterization of hybrid superconductor - quantum dot devices,which are likely to be used as building blocks of future conventional and topological quantum-information schemes. 

\acknowledgments
We acknowledge the fruitful discussions with Takis~Kontos, Pascu~C.~Moca and Gergely~Zar\'and. AP was supported by the National Research Development and Innovation Office of Hungary (NKFIH) Grants
%zarand
105149 and 
%asboth
124723, and the \'UNKP-17-4-III New National Excellence Program of the Ministry of Human Capacities of Hungary. This work was supported by NKFIH within the Quantum Technology National Excellence Program (Project No. 2017-1.2.1-NKP-2017- 00001), by the COST action NanoCoHybri CA16218, and QuantERA network 'SuperTop' (NN 127900).

\bibliographystyle{prsty}
\bibliography{scheriff}

\end{document}